# Teachers without borders: Beyond white chalk and black board

**An Essay**

Elochukwu A. Ukwandu  
Cardiff Metropolitan University, United Kingdom  
eaukwandu@cardiffmet.ac.uk

Robert Onyeneke  
Alex Ekwueme Federal University, Ndufu-Alike  
Robert.onyeneke@funai.edu.ng

## Abstract

An educator who is also known as a lecturer in the university system has three main areas of focus: learning (helping students to acquire knowledge, competence and virtue), research (developing new knowledge, breaking new grounds) and community service (applying the knowledge to real life situations to improve life and living conditions of the society). As the world's geographical boundaries keep getting redefined in the context of a global village, the constituency of teachers keeps getting redefined as well. This essay aims to address issues about modern constituent and platform of teachers in Nigeria for service delivery in the context of a globalised world. It also focuses on how to reach out to these new set of communities brought about by globalisation to remain relevant, effective and efficient alongside their perceived challenges and possible solutions in Nigerian context.

## 1. Introduction

The 21$^{st}$ century digital revolution has seen the world's geographical boundaries shrink significantly making it possible for information to be passed within seconds to any part of the world (Collins, A., & Halverson, R. (2018)). This has been made possible by internet originally meant to connect computers together known as International Network of Computers. But with the development of client-server software architecture, the possibility of interacting not only on computer level but on human level over the keyboard, computer monitor, and mouse has become possible and at ease. This communication has been made faster and swifter with the coming of global satellites for interconnections, pocket computers, digital assistants, mobile phones and other digital devices.

In the wake of this comes an addition of a new concept - electronic shortened as (e-) that has affected almost every facet of human endeavours known to man on earth such as in Commerce, Governance, Politics, and so on (Collins, A., & Halverson, R. (2018)). Teaching and Learning that traditionally had been discharged over the black board using white chalk in a classroom with students in attendance have also come under new redefinition and with wider jurisdiction as well. This new redefinition of teaching and learning in relation to how their new job descriptions are defined has become part of the focus of this essay. The reason we are exploring this new concept, jurisdiction and job description is to present it as it is to the Nigerian academic community as we know it. Moreso, to elicit knowledge, prospects, local challenges and perhaps help to reduce perceived lethargy, fear and misconceptions. The term university as conceived implies a universal concept in terms, definition and operation. and hence academics all over the world usually subject their research outputs to peer-review mechanism before it is published for public consumption. Suffice it to say also, that the concept of globalisation defined by knowledge-based economy comes with its attendant effect on the university system as well. This implies that the mode of teaching, research and community service as a way of having a university being heard where it is located has been redefined. Thus, community service, teaching and research have been completely globalised. The implication of this global phenomenon and definition forms the focus of this essay going forward.

## 2. The defined constituent of lecturers and their limitations in a globalised world

Traditionally, the impartation of knowledge is defined within the walls of the university of service in such a manner that is localised and distinctively controlled. That is to say, the lecturers depending on the level of contacts may know most of the students to a large extent as the job of knowledge transfer is done on a face-to-face mode daily. In the same vein, some lecturers often carry out their research and community services within the confine of the community they serve.

In line with the world as a global village and with the concept of academic freedom being enjoyed by a lecturer, s/he is deemed globally employable personnel. This implies that a lecturer can work anywhere in the world for economic, social, professional, or academic gains. But does teaching only in a university of employment/contract make this global phenomenon possible? What are the limitations of this current trend of teaching in a confined classroom in a globalised world? Realistically, a lecturer does research and publishes to be heard and be known as a professional in certain fields of endeavour. The implication is to be involved in not just services to the immediate community but to the global setting. Suffice it to say therefore that what a lecturer teaches ought to be to a large extent the product of his/her research findings, which have been tested in community services alongside necessary feedbacks both from the students and the communities being served with the knowledge generated.

The foregoing implies that if a lecturer is only confined to the classroom walls of his/her university of employment/contract locally, the teaching, research, services, and feedbacks will in no doubt be localised. This is a strong limiting force to a global phenomenon.

## 3. Redefining academic constituent in a universal concept

With the concept of globalisation, every facet of human endeavours is being redefined in line with global views and education is not left behind. It is therefore necessary to use this medium to make clear that in the 21$^{st}$ century, the concept of lecturer's classroom wall, audience and community are no longer within a localised jurisdiction rather a global setting. If the Nigerian educational system in higher education is to be seen, heard, recognised as playing an active role to the benefits of her local community and the world at large, it must have to key into this global trend. We will therefore use the following subheadings to establish the new concepts of classroom, publishing and teaching that have global reach and the attendant benefits to the academic community in Nigeria.

## 3.1. The virtual classroom

A virtual classroom, also known as Virtual Learning Environment (VLE) according to Techopedia (Techopedia, 2019), is a "teaching and learning environment that provides a platform for participant to interact, communicate, view and discuss presentations. They are also empowered to engage with learning resources while working in groups, all in an online setting. The medium is often through a video conferencing application that allows multiple users to be connected at the same time through the internet, which allows users from anywhere in the world to participate."

So, in addition to the traditional classroom, Nigerian universities should also add virtual learning environment where lecturers can engage, interact, share resources with students in teaching and learning all over the world, which is also known as Blended Learning Environment (Al-Samarraie, H., & Saeed, N. (2018)). In the same, as universities in developed world have changed the concept of Distance Learning (Moore, J. L., Dickson-Deane, C., & Galyen, K. (2011)), a situation where students used to attend to classes through correspondence to rather get them engaged through a virtual classroom setting. This involve the use of computer hardware, software and some peripheral devices like camera and video. Such classrooms setting now involve the use of video conferencing application such as Cisco WebEx, Microsoft Skype, Zoom, Google Meet, Microsoft Team, and others. Nigerian universities can use this concept to enhance their classroom administration for both regular and distance learning students.

## 3.2. New publishing, Teaching, Learning and Mentoring concepts

Beyond publishing research articles in peer-reviewed journals and conference proceedings, lecturers have added to it lecture notes in digital formats, videos of classroom sessions on newly developed knowledge and concepts, topical discussions using social media handles or Webinar as well as tutorials on perceived hard courses. The implication of this is that both the lecturer and university are being widely publicised to the world as search engines profiling on such activities make it easy for internet users to find them and use where necessary. In all, the lecturer is getting known as a voice not just within his/her academic jurisdiction but all over the world. This in turn, could lead to more economic, social, and professional gains as feedbacks are received from all over world thus making the lecturer more knowledgeable on what people say about his/her activities in terms of relevance and impartations.

In all, lecturers could mentor, supervise and examine seminars, projects, dissertations for upcoming academics wherever the student may reside. Furthermore, mentoring and supervisions can be done through video conferencing (such as Skype, Microsoft Teams, Slack, Google Meet, Zoom, etc.) and collaborative cloud storage document sharing system (such as in Dropbox, Box, and GitHub).

## 3.3. Teaching without borders

The idea of teaching without borders is conceived like that of Doctors Without Borders, where services are rendered outside teacher's/lecturer's primary place of work. The idea embodies the attributes that teaching is more of a vocation rather than a profession and hence teachers/lecturers are driven primarily by altruism, love and passion to build, develop and mould a society that works for all. In furtherance of this, the system conjures the concept of Academic Missionaries and Evangelists in carrying out this task. The platforms for service delivery are both online and offline, online through Video Conferencing applications using Virtual Classrooms, document sharing and collaboration through cloud-based storages, while offline, through face-to-face traditional pedagogy. The ratio is either 70% to 30% or 80% to 20% as may be agreed due to level of availability by the lecturer being engaged. The works of academic missionaries will be as conceived more of breaking new grounds by producing new knowledge, curriculum, concepts and up-to-date academic system as practised globally but with local contents that makes them relevant to local problems. The academic evangelists will be more of teaching the concepts so produced, as well as expanding it as proposed in relation to local problems. They can also be involved in making other online course material explicit to locals. So, in doing this, can as well teach through other online courses that are needed to keep the system up to date such as in using Stanford Coursera, MIT Open CourseWare, UdemyEdX and so on.

In a bid to define this further, the conception is driven by the belief that Nigeria's academic system in higher education is highly localised. The Authority News (The Authority New, 2020) is of the view that Nigerian universities are being "dangerously localised." To proffer solution to this, we have taken to this unusual approach, and we deemed it imperative as espoused with the focus that the university system will open to external influence by increasing the ratio of non-locals to locals in staff strength. This will be done by employing staff outside the geographical location of the university that have the qualities to turn things around, not unmindful that this might take time especially in state-owned institutions of higher learning that have meagre budget. Also, academic missionaries and evangelist are more likely to be accepted as they are not in any way competing with locals on issues of promotions, academic leadership, and politics, rather helping to improve teaching, learning and research of the local. In all, they are solely on part-time contract, living outside of Nigeria and driven by passion, love, and altruism to make Nigerian a better place for all.

**The merits of teaching without borders**

**Heterogeneity of ideas:** one of the main merits of teachers without borders as proposed is heterogeneity of ideas. We are of the view that part of the problems of Nigerian academic institutions that inhibit her growth in carrying out quality teaching, research and community services is the homogeneity of staff and students and hence ideas. There is therefore, a strongly perceived sense of heterogeneity in practising as proposed the concept of Teachers Without Borders (TWB). Others are cross-pollination of ideas; strong peer-review mechanism; cross-disciplinary, cultural, and geographical outputs, strong academic presence; good global rating, influence and affluence through problem solving; quality outputs and feedbacks. It will also lead to more economic, social, academic, and professional prospects for staff, students, and the university community.

## 4. The perceived challenges Nigerian teachers face in the new concept

### 4.1. Challenges relating to Virtual Classroom and Teaching without borders

Recent studies have reported several challenges facing the new concepts of virtual classroom and teaching without borders in higher education in Nigeria. We briefly review the challenges following broad thematic areas as follows:

#### 4.1.1. Technical Paradigm Shift

A transition to virtual learning amid traditional higher education culture, which represents a paradigm shift, introduces novel challenges worth considering. These challenges are mostly technical and important in nature due to their effects on the key stakeholders in the learning process; the educators (or faculty) and learners. Specifically, educators may be unprepared or antagonistic to the use of disruptive modern technologies for which they may be unskilled (Nsofor, Umeh, Ahmed, & Sani, 2014). The potential benefits that accrue due to the paradigm shift to virtual learning may be hindered by factors such as limited or lack of internet connectivity, virtual learning equipment such as computers and relevant software programmes (Eze, Chinedu-Eze, & Bello, 2018; Kayode, Alabi, Sofoluwe, & Oduwaiye, 2019). Access to Broadband Internet is low due to the prohibitive cost associated with Broadband Internet; hence, educators and learners have limited access due to limited incomes (Pandey & Tiwari, 2014). On the other hand, irregular electricity supply affects every sector of the Nigerian economy including the educational sector (Anene, Imam, & Odumuh, 2014; Eze *et al.*, 2018). Where ICT facilities are available, effective usage may be hampered by irregular power supply. Other potential challenges with such technical transition relates to low students' self-discipline, organisational skills and lack of decorum in keeping to virtual classroom instructions (Nsofor *et al.*, 2014).

#### 4.1.2. Educators

Another major challenge faced in the transition to the virtual classroom and teachers without borders is the low expertise with modern ICT tool. Due to the novelty of integration of virtual learning environments, educators potentially face a major pedagogical challenge. This includes transferring physical, face-to-face course materials to the virtual environment, while fostering students' engagement and effective learning (Kebritchi, Lipschuetz, & Santiague, 2017; Nsofor *et al.*, 2014). The transition to the virtual environment presents a major challenge in faculties where the educators are predominantly older than the Generations X and Y; older educators are trained in the traditional face-to-face medium and less likely to have been widely exposed to the intricacies of the online medium of teaching or less interested in transitioning to a virtual courses, thus will hamper smooth transitioning (Kebritchi *et al.*, 2017).

Considering content development and instructors, the contents of most online courses are predefined; hence this reduces the inputs of educators since they have to use them as given. The challenge in this context is mainly related to educators who played no role in designing the online courses; it precludes their ability to introduce new materials and modes of teaching from face-to-face settings (Nsofor *et al.*, 2014). The introduction of modern learning technologies is often associated with less attention to appropriate content design and budget to create a successful programme, while favouring concentration technology implementation. On the other hand, there are problems associated with managing and monitoring participant progress (Kebritchi *et al*., 2017). Unlike during in-person class learning where learners' progress can be easily monitored by the educators, monitoring, and managing learners' progress in the virtual environment is harder to implement especially where the Learning Management System are bundled software.

### 4.1.3. Learners

Students' learning expectations during a transition to a virtual learning environment may pose some challenges. Expectations such as prompt response to questions asked in online comments may prove frustrating due to delays in response where there is lack in personnel or long queues of questions, which is different when compared to in-person class learning. Without appropriate staff and student discipline needed in a virtual learning environment, students may not take homework deadlines seriously in a virtual learning environment compared with in-person class learning (Nsofor *et al*., 2014). On the technology front, many Nigerian higher institutions may not have some basic computer education programmes for fresh student, thus creating room for computerphobia, and with the inability of some of these students to engage in part-time employment, the burden of providing Laptops, Smart phones or Personal Digital Assistance that will support online learning will now be on their parents. Furthermore, many lack access to reliable Internet access that are capable of providing video conferencing, calls and cloud sharing, which further limits the effectiveness of the deployment of a virtual learning environment (Anene *et al*., 2014; Eze *et al*., 2018; Olutola, 2015).

## 4.2. New Publishing

The creation and dissemination of scientific knowledge in the Nigerian educational system is facing various challenges. These challenges include, but are not limited to the high financial constraints associated with publishing in, and running high-quality open access journals (Ajao & Ugwu, 2011). Furthermore, inexperienced educators in higher education face the problems of journal assessment such as journal indexing and prevalence of predatory journals (Ajao & Ugwu, 2011). The prevalence of predatory journals is fostered by lack of knowledge on journal assessment metrics and indexes. These factors may result in damaged external reputation of educators and their host institutions (Ajao & Ugwu, 2011; Richtig, Berger, Lange-Asschenfeldt, Aberer, & Richtig, 2018).

## 5. Possible solutions

Equipping virtual classrooms with e-learning technologies, training of lecturers and improving the internet connectivity and power supply should be on the priority list of higher education administrators for a swift and seamless transition to online teaching and learning in the country. Furthermore, financing plans for educators and learners should be created to fund access and ownership of ICT equipment such as laptops and essential software programmes to enhance smooth e-teaching and e-learning. Other stakeholders such as the industry, civil society groups and donor organizations should contribute towards equipping the learning centres, constant training of staff, maintenance and security of ICT equipment and facilities(Anene *et al*., 2014; Eze *et al*., 2018; Olutola, 2015). Universities' e-libraries should be well-equipped and encouraged to subscribe to online resources (journals, books, monographs and conference proceedings). Students and lecturers should be given access to such online resources to enhance effective e-teaching and e-learning. Redesign the training of educators to include induction and mentoring during pre-service and in-service periods. There should be periodic systematic review of educator training curricula, establishing schools of education for continuous professional education and introducing administrative academic qualifications (Nsofor *et al.*, 2014).

## 6. Summary and Conclusions

This essay addressed the emerging role of the university lecturers or educators in modern times. It proposes a new concept of teaching, learning, and research, for education stakeholders in Nigeria. Traditionally, lecturers (educators) impart knowledge to learners (students) in a classroom setting in such a manner to avoid external distractions. However, modern concepts have evolved to augment the traditional modes of learning; these concepts include, but are not limited to, virtual learning or classroom, and proposed Teachers Without Borders. Virtual classroom provides a platform for participants to interact, teach and learn using modern information communication technologies such as internet-connected computers, while TWB is a new concept conceived to help open up Nigerian education space to external influence where by Nigeria can play an active role both as benefactors and beneficiaries to enhance its teaching, learning, research and community service in a globalised world.